\documentclass[pdflatex,sn-mathphys-num]{sn-jnl}

\usepackage{graphicx}%
\usepackage{multirow}%
\usepackage{amsmath,amssymb,amsfonts}%
\usepackage{amsthm}%
\usepackage{mathrsfs}%
\usepackage[title]{appendix}%
\usepackage{xcolor}%
\usepackage{textcomp}%
\usepackage{manyfoot}%
\usepackage{booktabs}%
\usepackage{algorithm}%
\usepackage{algorithmicx}%
\usepackage{algpseudocode}%
\usepackage{listings}%
\usepackage[mathlines]{lineno}
\usepackage[capitalize]{cleveref}
\usepackage{pdfpages} 
\usepackage{pgffor} 
\usepackage{xr} 

\usepackage{pdfpages} 
\usepackage{pgffor} 
\usepackage{xr} 

\makeatletter
\AtBeginDocument{\let\LS@rot\@undefined}
\makeatother

\def\supplementfilename{supplementaryMaterial/supplementaryMaterial}

\pdfximage{\supplementfilename.pdf}
\def\numbersupplementpages{\the\pdflastximagepages}

\newif\ifarXiv
\arXivtrue 

\Crefformat{figure}{#2Fig.~#1#3}
\Crefmultiformat{figure}{Figs.~#2#1#3}{ and~#2#1#3}{, #2#1#3}{ and~#2#1#3}
\Crefformat{equation}{#2Eq.~#1#3}

\graphicspath{{Figures/}}
\newenvironment{ruledtabular}{}{}

\begin{document}


\theoremstyle{thmstyleone}%
\newtheorem{theorem}{Theorem}
\newtheorem{proposition}[theorem]{Proposition}%

\theoremstyle{thmstyletwo}%
\newtheorem{example}{Example}%
\newtheorem{remark}{Remark}%

\theoremstyle{thmstylethree}%
\newtheorem{definition}{Definition}%

\raggedbottom

\title[Article Title]{Optomechanical vector sensing of new forces at 6 micron separation
}


\author*[1]{\fnm{Gautam} \sur{Venugopalan}}\email{gautamve@stanford.edu}

\author*[1]{\fnm{Clarke A.} \sur{Hardy}}\email{cahardy@stanford.edu}

\author[1]{\fnm{Kenneth} \sur{Kohn}}

\author[1]{\fnm{Yuqi} \sur{Zhu}}

\author[1]{\fnm{Charles P.} \sur{Blakemore}}

\author[1]{\fnm{Alexander} \sur{Fieguth}}

\author[1]{\fnm{Jacqueline} \sur{Huang}}

\author[1]{\fnm{Chengjie} \sur{Jia}}

\author[1]{\fnm{Meimei} \sur{Liu}}

\author[1]{\fnm{Lorenzo} \sur{Magrini}}

\author[1]{\fnm{Nadav} \sur{Priel}}

\author[2]{\fnm{Zhengruilong} \sur{Wang}}

\author[1,3]{\fnm{Giorgio} \sur{Gratta}}


\affil*[1]{\orgdiv{Physics Department}, \orgname{Stanford University}, \orgaddress{ \city{Stanford}, \postcode{94305}, \state{CA}, \country{USA}}}

\affil[2]{\orgdiv{Applied Physics Department}, \orgname{Stanford University}, \orgaddress{\city{Stanford}, \postcode{94305}, \state{CA}, \country{USA}}}

\affil[3]{\orgdiv{Hansen Experimental Physics Lab}, \orgname{Stanford University}, \orgaddress{\city{Stanford}, \postcode{94305}, \state{CA}, \country{USA}}}


\abstract{The search for new gravity-like interactions at the sub-millimeter scale is a compelling area of research, with important implications for the understanding of classical gravity and its connections with quantum physics.
    We report improved constraints on Yukawa-type interactions in the $10\,\mathrm{\mu m}$ regime using optically levitated dielectric microspheres as test masses. The search is performed, for the first time, sensing multiple spatial components of the force vector, and with sensitivity improved by a factor of $\sim 100$ with respect to previous measurements using the same technique.  The resulting upper limit on the strength of a hypothetical new force is $10^7$ at a Yukawa range $\lambda\simeq 5\;\mu$m and close to $10^6$ for $\lambda \gtrsim 10\;\mu$m. This result also advances our efforts to measure gravitational effects using micrometer-size objects, with important implications for embryonic ideas to investigate the quantum nature of gravity.}




\maketitle

\section{Introduction}\label{sec1}

At large distances, gravity is the most familiar and best characterized of the four fundamental interactions. The motion of bodies in the Solar System has provided a platform for exquisite tests of Newtonian gravity \cite{merkowitz_tests_2010}, while precision tests of General Relativity validate its use to describe phenomena also in the strong gravity regime ~\cite{the_ligo_scientific_collaboration_and_the_virgo_collaboration_tests_2019,ligo_scientific_collaboration_and_virgo_collaboration_tests_2021}. On short length scales, however, the nature of the gravitational force is far less well explored. Indeed, it has never been directly measured for face-to-face separations between test masses smaller than $52\,\mathrm{\mu m}$~\cite{lee_new_2020}, and constraints on modifications to the inverse square law (ISL) weaken by orders of magnitude at distances shorter than $10~\mu$m~\cite{adelberger_tests_2003}. Yet, the Standard Model (SM) of fundamental interactions implicitly assumes that the ISL applies unchanged all the way down to the Planck scale~\cite{arkani-hamed_universes_2000}.  Models beyond the SM include a rich landscape of predictions that would manifest as modifications to gravity on short length scales, including the existence of extra dimensions or of new light bosons~\cite{antoniadis_new_1998,sundrum_fat_2004,montero_dark_2023,safronova_search_2018,murata_review_2015}. This broad class of new physics is typically parameterized by adding a Yukawa term to the gravitational potential between two point masses $m_1$ and $m_2$,
\begin{equation}
    V(r) = -\frac{Gm_1m_2}{r}(1+\alpha e^{-r/\lambda}),
    \label{eq:yukawa}
\end{equation}

\noindent where $G$ is Newton's gravitational constant, $r$ is the distance between the point masses, and $\alpha$ and $\lambda$ are the strength and length scale, respectively, of the Yukawa modification.  The most stringent constraints on modifications to the ISL of this type have so far been established by measuring responses of mechanical oscillators, often torsion pendulums~\cite{chen_stronger_2016,lee_new_2020,tan_improvement_2020,sushkov_new_2011,geraci_improved_2008}. These techniques search for variations of one component of a temporally modulated force vector at one or more harmonics of the modulation frequency. In contrast, the technique presented here uses the time-dependent behavior of the full 3-dimensional force vector to search for new interactions. By conducting the search for a potential signal across multiple spatial dimensions and harmonics, the unique spectral fingerprint of a potential interaction can be exploited, especially to constrain and confirm a possible discovery. 

\section{Experimental Setup}\label{sec2}

\begin{figure}[!t]
    \includegraphics[width=1\columnwidth]{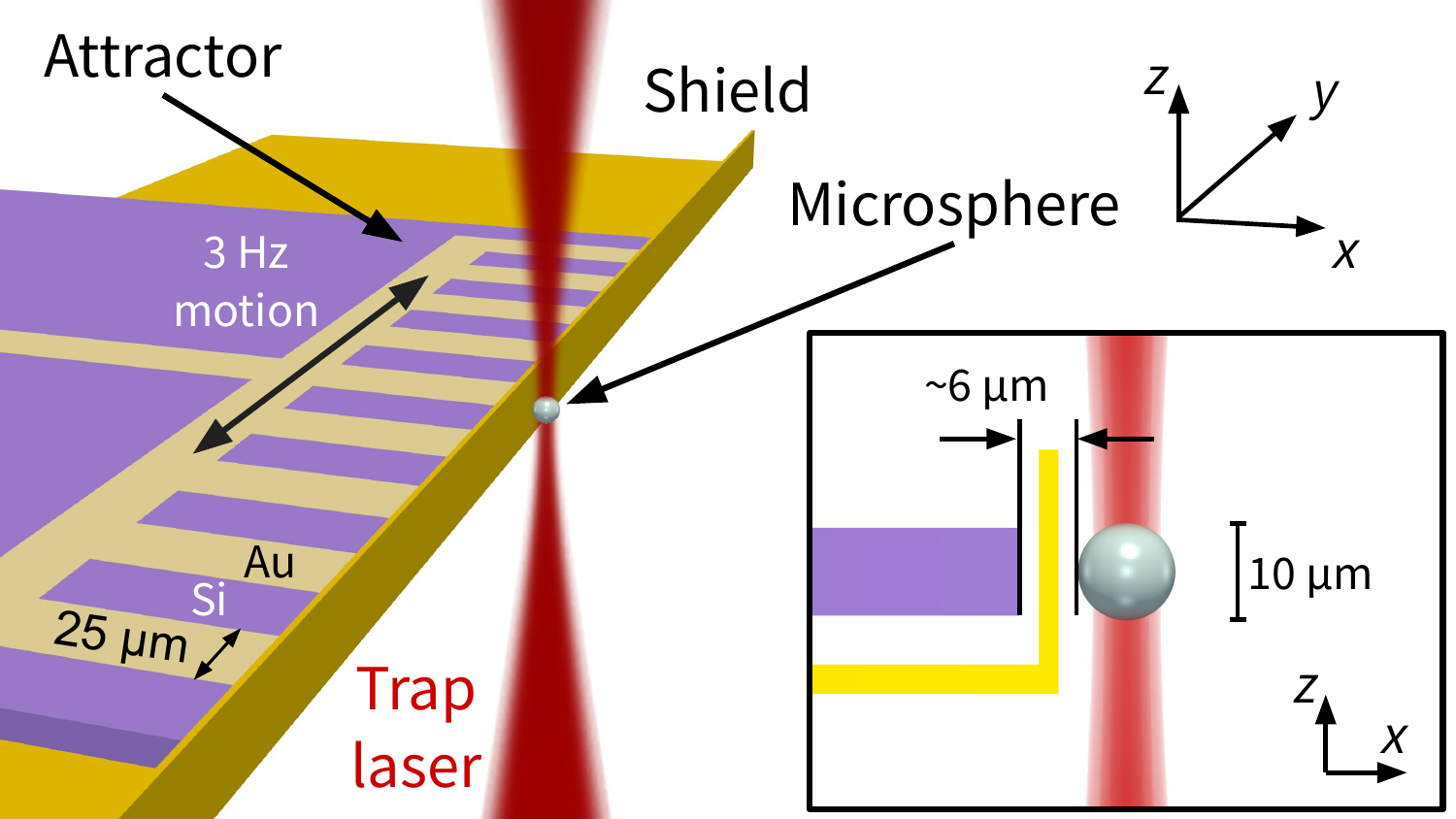}
    \includegraphics[width=1\columnwidth]{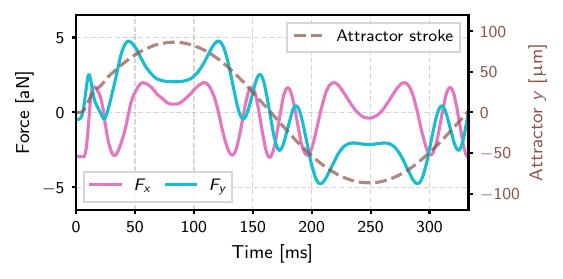}
    \caption{Top: diagram of the experimental setup also defining the coordinate system used and providing the most important physical dimensions. A MS is held at the focus of the trapping beam while the density-patterned attractor oscillates by $170\,\mathrm{\mu m _{pp}}$ at $f_0=3\,\mathrm{Hz}$ behind the stationary shield. The inset at the top shows a side view with the MS diameter and the face-to-face separation between the MS and attractor identified. The Platinum Black coating on the attractor, which hides the underlying density pattern, has been omitted in this diagram. Bottom: time evolution of the force exerted on a $10~\mu\mathrm{m}$ MS by a hypothetical Yukawa interaction with the attractor ($\alpha=10^6$, $\lambda=10~\mu\mathrm{m}$). The $z$ component of the force is not shown as it is only present when a vertical offset is introduced between the MS and the attractor. The secondary $y$ axis shows the position of the center of the attractor as it is driven through one full cycle.}
    \label{fig:expsetup}
\end{figure}

The sensor at the heart of this work is an optically levitated silica microsphere (MS)~\cite{noauthor_microparticles_nodate} trapped in a single-beam, vertically oriented optical tweezer. An early generation of the setup is described in~\cite{kawasaki_high_2020}. The experimental geometry is illustrated in \Cref{fig:expsetup}. A $1064\,\mathrm{nm}$ laser beam is focused with an off-axis parabolic mirror to a waist of $\sim 3.5\,\mathrm{\mu m}$ (Numerical Aperture $\mathrm{NA}\sim 0.1$, Rayleigh range $z_{\mathrm{R}} \sim 30\,\mathrm{\mu m}$). The MS is trapped near the focus, allowing the closest distance of approach of a source mass (``attractor") with minimal interaction with the trapping beam. The stabilization of laser intensity and angular jitter by real-time feedback allows the MS to be isolated from extraneous disturbances at high vacuum ($\sim10^{-7}\,\mathrm{hPa}$). The light scattered off the MS in the forward direction is re-collimated by a second off-axis parabolic mirror and used to reconstruct the $x$ and $y$ positions of the MS~\cite{maurer_quantum_2023} using a quadrant photodiode (QPD). The phase of the retroreflected field from the MS is interferometrically sensed to reconstruct the $z$ coordinate. 

The trap is surrounded by six electrodes (three pairs, one for each Cartesian axis, separated by $\sim 8\,\mathrm{mm}$). The electrodes are spray-coated with AquaDAG \cite{acheson_aquadag_1907} to reduce reflections of stray laser light. The monopole electric charge on the MS can be controlled with single-electron resolution \cite{moore_millicharge_2014, blakemore_3d_2019}, allowing for a bias on the electrodes to exert a known force on the MS, thereby enabling the calibration of the optical position readout to force. The entire readout chain is found to be linear up to $10^{-13}\,\mathrm{N}$~\cite{blakemore_3d_2019}, well beyond the forces of interest here.

A density-patterned attractor~\cite{wang_density_2017} acts as a source for interactions coupling to mass. Gold and silicon are used as the two materials generating a contrast in density, $\rho$, with $\rho_{\mathrm{Au}} \sim 8\rho_{\mathrm{Si}}$. By scanning the position of the attractor along $y$ in close proximity to the MS at a frequency $f_0=3~\mathrm{Hz}$, the MS is subject to a characteristic force with a distinctive spectral fingerprint in each Cartesian axis that is unlikely to be mimicked by backgrounds. As electromagnetic forces constitute an important source of background, several design choices were made to mitigate their contribution. To prevent time-varying magnetic fields in the trap, the attractor is fabricated from materials that are not permanently magnetizable. The MS can be made to have zero overall charge, though dipole and higher order moments persist and can couple to time-varying electric field gradients as the attractor is scanned. In order to mitigate this effect, a stationary ``shield" (a microfabricated silicon fence structure, conformally coated with $\sim 100\,\mathrm{nm}$ gold) is positioned between the MS and the scanning attractor.  In addition, using the six electrodes surrounding the trap, a rotating electric field can be applied to the dipole moment inducing the rotation of the MS, as explained in Section~\ref{sec4}.

The first constraints on modifications to gravity using a levitated force sensor were reported in~\cite{blakemore_search_2021}. The present iteration of the experiment incorporates several improvements over the first. The amount of stray light scattering off the moving attractor and resulting in a background on the detectors was reduced $\gtrsim100-$fold. This was primarily achieved by adding a spatial filter at the focus of a downstream telescope, and by coating the attractor with a $\sim3\,\mathrm{\mu m}$-thick layer of Platinum Black~\cite{venugopalan_platinum_2024}, with a reflectivity at $1064\,\mathrm{nm}$ that is $\lesssim 1\%$ that of the original gold. In addition, various parts of the vacuum chamber were stiffened to mitigate the effect of mechanical vibrations, and a series of accelerometers and microphones were added to sense the seismic and acoustic environment close to the setup. Optimal (Wiener) filters~\cite{allen_automatic_1999,vajente_data_2020} were then constructed to subtract environmental disturbances coherently sensed by the MS and the environmental sensors. Lastly, the noise in the force measurement was improved by over a factor of 10 along the $x$ and $y$ axes, and $\sim 5\times$ along the $z$-axis. This improvement was achieved primarily with better stabilization of the laser intensity delivered to the trapping region (relative intensity noise, $\mathrm{RIN}\lesssim 5\times 10^{-7}/\sqrt{\mathrm{Hz}}$). With these upgrades, the forward-scatter detection system has a displacement sensitivity of $\sim 10^{-11}~\mathrm{m/\sqrt{Hz}}$, which translates to a force sensitivity of $\sim10^{-17}~\mathrm{N/\sqrt{Hz}}$ for the typical trap spring constant of $\sim10^{-6}~\mathrm{N/m}$.

\section{Analysis}\label{sec4}

The results presented in this work were obtained from measurements of three MSs, one with nominal diameter $7.56~\mu\mathrm{m}$ and the other two $9.98~\mu\mathrm{m}$. Their masses are measured in situ, following the procedure in ~\cite{blakemore_2019_mass}. A detailed list of the experimental conditions under which data was collected is available in~\cite{supp}. The search for new interactions is done by determining the value of $\alpha$ that provides the best fit between signal templates and the measured forces across multiple dimensions and harmonics. The fit includes both the magnitude of the force component and its phase with respect to the oscillation of the attractor. Signal templates are computed for various values of $\lambda$ using a finite-element model of the geometry of the attractor and MS over a spatial grid of the trap volume. An example template in the time domain is shown at the bottom of \Cref{fig:expsetup}. The path traversed by the attractor during each measurement is recorded and used to construct the signal template for that particular trajectory. Both positive and negative values of $\alpha$ are considered, enabling a search for attractive and repulsive modifications to the Newtonian gravitational potential. The data presented here were collected with negligible offset between the attractor and MS centers-of-mass in the $z$ direction, maximizing the signal along the $x$ and $y$ directions, at the expense of that in $z$.  

\begin{figure}
    \includegraphics[width=\columnwidth]{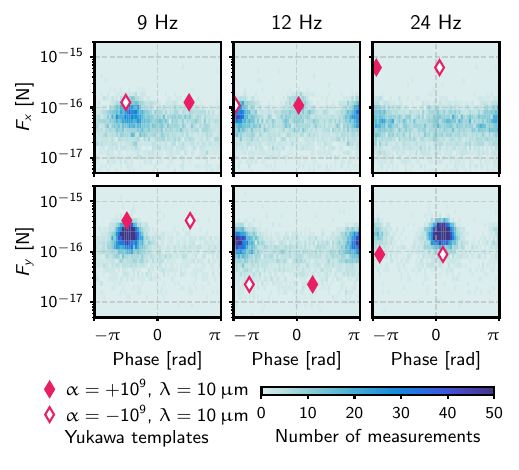}
    \caption{\label{fig:polar}Measurements of the $x$ and $y$ components of the force vector at the three harmonics with the largest observed power. The filled (unfilled) diamonds show the expected signal magnitudes and phases for an attractive (repulsive) Yukawa interaction with $\alpha$ and $\lambda$ shown in the legend. The measurements are not consistent with either signal template across all harmonics.}
    \end{figure}

While some power is present above the noise level at some of the harmonics of the 3~Hz scanning frequency, the relative distribution of such power does not match the Yukawa signal template and therefore the experiment found no evidence of a new interaction. To illustrate some features of the background, \Cref{fig:polar} shows the magnitude and phase of the effective forces measured for  three harmonics with the largest power above the noise. The magnitudes and phases expected for true attractive and repulsive interactions parametrized by \Cref{eq:yukawa} are also shown, allowing for a direct comparison between the measurements and the signal template. Three sources of background are identified: mechanical vibrations, electromagnetic coupling, and scattering of stray light. All three of them substantially improved with respect to the previously reported results.

\begin{figure*}
\includegraphics[width=\textwidth]{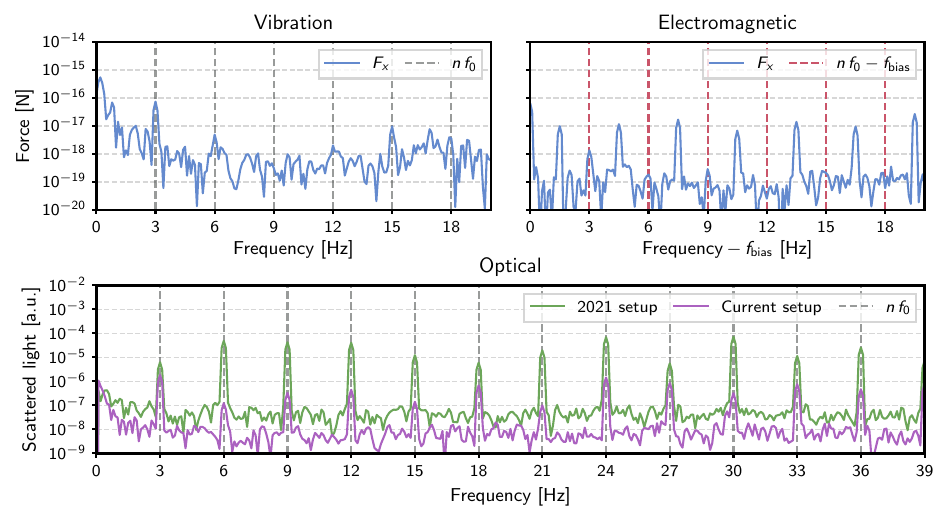}
\caption{\label{fig:backgrounds}Force spectra showing the measured backgrounds from three sources (from three different MSs). Top left: vibration backgrounds measured when the  attractor is driven in a retracted position; top right: in situ measurement of electromagnetic backgrounds at the sidebands of the bias modulation frequency (note the different centering of the frequency scale). The measured forces have been projected along the phase of the signal template to reflect only the component limiting sensitivity. Bottom: scattered light backgrounds measured by the QPD null stream in the 2021 setup~\cite{blakemore_search_2021} and the current setup, after having added the spatial filter and coated the attractor with Platinum Black.}
\end{figure*}

The mechanical coupling of vibrations from the oscillating attractor stage may cause relative motion between the MS, the beam, and the rest of the hardware. This contribution is measured by retracting the attractor along $x$ by a large distance ($\sim 5\,\mathrm{mm}$) from the nominal measurement position. The left panel of Fig.~\ref{fig:backgrounds} shows the measured spectrum under these conditions, demonstrating that vibration backgrounds are present predominantly at the fundamental frequency. Excluding the fundamental frequency from the analysis renders this background negligible with minimal loss in signal power.

Interactions between the electric dipole moment (EDM) of the MS and electric field gradients~\cite{priel_dipole_2022} from the attractor were already identified as another possible source of backgrounds. While the shield attenuates direct coupling, substantially better alignment and better control of the MS spinning have reduced this background to a sub-dominant strength. The effects of a contact potential difference between the attractor and shield could still produce field gradients at the location of the MS which would change as the relative position of the attractor and shield does. Measurements of the EDM of MSs using the technique described in~\cite{blakemore_librational_2022} give a typical EDM of $\mathcal{O}(100\, e\cdot\mu m)$. 

A rotating ($f_\mathrm{rot}\sim 200\,\mathrm{kHz}$) electric field, applied using the electrodes surrounding the trap, is used to confine the EDM of the MS to a plane that minimizes backgrounds. Empirically, drifts in the out-of-plane component of the EDM are typically $\lesssim 10\%$ over the course of a measurement. An initial investigation of the remaining electromagnetic background was carried out in situ by applying an amplitude modulation to the bias between the attractor and shield at a frequency $f_\mathrm{bias}\gg f_0$ ($f_\mathrm{bias}\ll f_{\mathrm{rot}}$). Forces coupling to the EDM, which normally appear as backgrounds at the harmonics of the attractor's mechanical drive $n f_0$, would manifest as sidebands at $f_\mathrm{bias}\pm n f_0$. A Yukawa interaction is not expected to couple to the EDM, ensuring that these sidebands remain clean background witness channels, although potential effects of an induced dipole in the MS requires more investigation. The right panel of Fig.~\ref{fig:backgrounds} shows the typical fingerprint obtained by this process (with a $150\,\mathrm{mV_{pp}}$ bias modulation and  $f_\mathrm{bias}=40.5~\mathrm{Hz}$), indicating that EDM backgrounds appear to be sub-dominant with the current force sensitivity.

 The dominant background currently originates from stray light from the trapping beam scattering off the scanning attractor and being imaged on the QPD. While great effort is taken to ensure good spatial quality of the trapping beam around the focus, ignoring the shield, it is estimated that $\mathcal{O}(100\,\mathrm{ppm})$ of light from the Gaussian ($\mathrm{TEM_{00}}$) trapping beam exists at the location of the attractor. In addition, it is plausible that a small fraction of light scatters off the AquaDAG-coated electrodes surrounding the trap volume and into the $x-y$ ($z$) QPD (photodiode) detection systems, modulated by the reciprocating attractor. Investigations of this background have qualitatively confirmed several expected properties: (i) it is present in some form when the attractor scans with its nominal cycle but without a MS in the trap, (ii) it decreases as the attractor is retracted away from the MS, (iii) it increases as the attractor moves vertically away from the focus, cutting into more of the beam, and (iv) it was drastically reduced by the combination of using an attractor coated by absorbent Platinum Black and adding a spatial filter ($50~\mu\mathrm{m}$ diameter Acktar~\cite{Acktar:2024} coated aperture) to the downstream optics to select the mode containing information about MS motion. The first point allows for some aspects of this background to be studied in the absence of a MS. However, the Mie-scattering distribution from the MS significantly alters the light scattering pattern, preventing this technique from being used for quantitative background modeling. In addition, the background varies in time, making its effective subtraction more challenging. The QPD used to sense forward-scattered light from the MS consists of 4 separate sectors arranged in a square, from which four linearly-independent combinations can be constructed. Three of these are constructed to report the $x$ and $y$ positions of the MS, and the sum of light incident on the QPD. The fourth remaining linear combination is interpreted as a ``null stream," which is sensitive to random spatially fluctuating light fields, but empirically verified to be insensitive to a true force transduced by the MS. This construction of data streams from the QPD quadrants (described in detail in Sec 8.4.1 of \cite{hardy_search_2025}) was used only for estimating the contribution of stray light backgrounds. In the search for Yukawa interactions, the standard position data streams, given by Eq 8.1 of \cite{hardy_search_2025} were used. As shown in the bottom panel of Fig.~\ref{fig:backgrounds}, the null stream shows substantial background reduction when a Platinum Black coated attractor is used in combination with the spatial filter, mitigating the effects of stray light.

As the measured forces are not consistent with the signature of a new Yukawa interaction, we report upper limits on the strength parameter $\alpha$ for a range of $\lambda$ values. A likelihood function is defined as

\begin{equation}
\begin{alignedat}{2}
    \mathcal{L}(\alpha|\lambda) =  \prod_{i,j,k} \frac{1}{2\pi\sigma_{ijk}^2}
     & \times \exp\left\{-\frac{\left[\Re\left(F_{ijk} - \alpha\,\tau_{ijk}(\lambda)\right)\right]^2}{2\sigma_{ijk}^2}\right\} \\
    & \times \exp\left\{-\frac{\left[\Im\left(F_{ijk} - \alpha\,\tau_{ijk}(\lambda)\right)\right]^2}{2\sigma_{ijk}^2}\right\},
\end{alignedat}
\end{equation}
where $F_{ijk}$ is the discrete Fourier transform of the $i^\mathrm{th}$ measurement at harmonic $j$ of the force component along axis $k$, $\tau_{ijk}(\lambda)$ is the corresponding signal template, and $\sigma_{ijk}$ is the RMS noise for the relevant measurement, harmonic, and axis combination, measured with the attractor stationary at its nominal equilibrium position. Due to observed backgrounds which do not exhibit a spectral fingerprint consistent with \Cref{eq:yukawa}, but which nonetheless vary significantly in their magnitude and phase, the choice of which harmonics to include in the analysis can affect the resulting limit. Including too many harmonics simply weakens the results by including channels with large observed backgrounds and little signal power, worsening the fit to $\alpha$. Using too few harmonics results in discarding a large fraction of the available signal power, and increases the sensitivity to the particular selection of harmonics used, and thus single-frequency systematic effects. To strike a balance, for each polarity of $\alpha$, we use the combination of six harmonics that maximizes the goodness of fit, quantified by the $\chi^2$. This criteria conservatively selects the harmonics that most mimic the spectral behavior of a potential signal while ensuring that $>50\%$ of the spectral power of such a signal is captured. With the likelihood construction described above, the profile likelihood ratio reduces to a simple parabolic form,
\begin{equation}
    -2\log{\frac{\mathcal{L}(\alpha|\lambda)}{\mathcal{L}(\hat{\alpha}|\lambda)}} = \frac{(\alpha - \hat{\alpha})^2}{\sigma_\mathrm{stat}^2},
\end{equation}
where $\hat{\alpha}$ is the maximum likelihood estimator for $\alpha$ and $\sigma_\mathrm{stat}$ describes the statistical uncertainty on $\hat{\alpha}$. Constraints can then be placed on the strength parameter $\alpha$ using a test statistic for upper limits, $q_\alpha$, defined as
\begin{equation}
    q_{\alpha} = 
    \begin{cases}
        \frac{(\alpha - \hat{\alpha})^2}{\sigma_\alpha^2} &|\alpha|\geq|\hat{\alpha}| \\
        0 &|\alpha|<|\hat{\alpha}|
    \end{cases},
\end{equation}
where $\sigma_\alpha^2 = \sigma_\mathrm{stat}^2 + \sigma_\mathrm{sys}^2$ is the total variance including both statistical and systematic uncertainties. The systematic uncertainties considered, along with their effect on the estimate of $\alpha$, are summarized in Table \ref{tab:systematics}.
\begin{table}[b]

\caption{\label{tab:systematics}%
Systematic uncertainties in measured parameters and their contribution to the uncertainty on $\alpha$ at $\lambda=10~\mathrm{\mu m}$.}
\begin{ruledtabular}
\begin{tabular}{lcr}
\textrm{Parameter, $p$}&
\textrm{$\Delta p$}&
\textrm{$\Delta \alpha / \alpha$}\\
\hline
Transfer function amplitude & 10\% & 0.1\\
MS mass & $20~\mathrm{pg}$ & $0.04$ \\
MS $x$ position & $0.7~\mu\mathrm{m}$ & $0.1$ \\
MS $y$ position & $1~\mu\mathrm{m}$ & $0.01$ \\
MS $z$ position & $0.5~\mu\mathrm{m}$ & $0.01$ \\
Attractor thickness & $1~\mu\mathrm{m}$ & $0.1$ \\
Platinum Black thickness & $1~\mu\mathrm{m}$ & $0.01$\\
Platinum Black density & $10~\mathrm{g/cm^3}$ & $0.01$ \\

\end{tabular}
\end{ruledtabular}
\end{table}
The 95\% confidence level limit on $\alpha$ for each sign is determined using the asymptotic behavior of the test statistic given by Wilks' theorem \cite{wilks_1938}. This statistical procedure has been tested using synthetic datasets created by injecting (in post-processing) a signal into data collected with the attractor stationary, ensuring that a true signal can be reconstructed accurately. Unmodeled backgrounds present in both polarities across multiple harmonics can add to or subtract from a real signal in the data, potentially resulting in misreported statistical coverage. To ensure robustness against this, synthetic signals were combined with datasets containing the measured backgrounds. Multiple statistical frameworks were then used to reconstruct the injected signals, and the framework demonstrating the most accurate coverage was selected.

\section{Results}\label{sec6}

\begin{figure}[!t]
    \includegraphics[width=1\columnwidth]{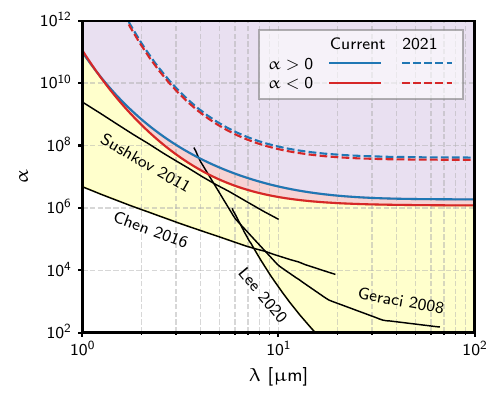}
    \caption{Constraints on the $\alpha-\lambda$ parameter space obtained in this work and the prior, 2021 result \cite{blakemore_search_2021}. The yellow shaded region indicates the region of parameter space excluded by other experiments~\cite{chen_stronger_2016,lee_new_2020,sushkov_new_2011,geraci_improved_2008}.}
    \label{fig:limits}
\end{figure}

Limits on $\alpha$ for new attractive and repulsive interactions for $\lambda$ ranging from $1-100~\mu\mathrm{m}$ were computed for each of the datasets~\cite{supp} considered. The strongest constraints among these are reported in Fig. \ref{fig:limits}. These new constraints represent an improvement of a factor of $\sim50$ over previous results for length scales of $\lambda>10~\mu\mathrm{m}$, and $\gtrsim100$ at $\lambda=2~\mu\mathrm{m}$. Moreover, they are the first to use reconstruction of multiple spatial components of the force vector as a function of time, hence exploiting the unique signature of a new force and providing a robust pathway for discovery. For example, a $z$ offset in either direction (c.f. the negligible offset used to collect data analyzed in this work) could be used to verify that a signal with the appropriate signature is transferred to this degree of freedom in case of a discovery.

The sensitivity of this search is limited by backgrounds primarily due to stray laser light scattering off the attractor, creating modulations of the optical power on the QPD at harmonics of the attractor drive frequency. While substantial progress has been made in reducing this and other backgrounds, potential avenues for further improvement have been identified. Work is underway to implement a low-noise, high dynamic range and frame rate $100$-pixel sensor, to be used for offline discrimination between true MS motion and scattered light effects. An alternative attractor geometry, consisting of a $30\,\mathrm{\mu m}$ thick, $1.5\,\mathrm{mm}$ diameter rotating disk with azimuthal density modulations~\cite{hough_novel_2022}, has also been prepared for potential use in the next experimental run. Better free space optics upstream of the trap is expected to produce more stable conditions and backgrounds, specifically addressing slow drifts of the trap focus which appear correlated with small variations in the refractive index of air driven by atmospheric pressure fluctuations. Beyond these immediate steps, a completely enclosed attractor should prevent any direct coupling between its motion and the beam, vastly attenuating optical backgrounds. Design of such an attractor system is a technical challenge to which future engineering efforts will be dedicated. Smaller and more constant stray light effects will facilitate the investigation of the remaining background, using some of the techniques described here.



\section{Conclusion}\label{sec13}

In this paper we have described the use of optically levitated MSs as force sensors capable of probing new Yukawa interactions that couple to mass at micron-scale separations. In addition to complementing existing techniques, this platform uses the fully-reconstructed force vector to search for the complex temporal signature that would arise from such interactions, a feature that may become crucial should future work result in a discovery. Techniques to probe the main sources of background limiting the experiment, as well as mitigation strategies, have been identified. Apart from the direct physics goal of the experiment, this work may be relevant to other uses of levitated optomechanics for fundamental physics \cite{moore_searching_2021}, including searches for dark matter interactions \cite{monteiro_search_2020,kilian_dark_2024}, sterile neutrino emission in weak decays \cite{carney_searches_2023}, or tests of the neutrality of matter \cite{priel_dipole_2022}. In addition, this work demonstrates the operation of a setup with microscopic objects levitated a few micrometers away from mechanical structures, while maintaining stable conditions, exploring backgrounds, and performing precision metrology. These are among the challenges that will need to be addressed towards experimental tests of the quantum nature of gravity~\cite{bose_spin_2017, gonzalez-ballestero_2021,millen_optomechanics_2020}.

\backmatter

\bmhead{Supplementary information}

Details of experimental configurations under which data was collected is included in the Supplementary Information.

\bmhead{Acknowledgements}
We gratefully acknowledge early contributions of Akio Kawasaki (AIST, Japan), Alex Rider (Scitech, Boulder CO), and Qidong Wang (IME-CAS, Beijing). We thank Giovanni Ferraro and Emiliano Fratini (University of Florence, Italy) for help in understanding some characteristics of St\" ober microspheres, and David Moore (Yale) for his feedback on the manuscript.



\section*{Funding information}

This work was supported by NSF grant number 2406999, ONR grant number N000142312600, and the Heising-Simons Foundation. Part of the work was performed at the Stanford Nano Shared Facilities (SNSF) which is supported by the NSF under award ECCS-2026822.  

\section*{Data availability}
The datasets used and/or analysed during the current study available from the corresponding author on reasonable request.

\section*{Author contributions}
G.V., A.F. and G.G. led the conceptualization, design and implementation of the experimental setup used in this work. Y.Z., C.B., J.H., M.L, and L.M. designed and constructed important subsystems. C.H. and G.V. led the data analysis, and K.K. developed numerical simulations used to validate the results. All authors reviewed the manuscript.

\bibliography{main.bib}

\ifarXiv
    

\section*{Supplementary material}

# Supplementary Material

## Summary of measurement configurations

The table below summarizes the measurement configurations used to produce the results in the main text. Each measurement consisted of 10 s of data collected at a sampling rate of 5 kHz. The quoted integration times are not limited by stability of the experiment - rather, the presence of backgrounds mean that there is no marginal benefit to integrating for longer times. Compared to an earlier iteration of the experiment [21], backgrounds and noise have been substantially improved. In addition, two important improvements result in a larger signal: (i) decrease of the $x-$ separation between attractor and microsphere (MS) by $\sim 4\,\mu$m (30%), and (ii) the use of heavier MSs. In the table below, MS diameters are nominal ones, verified using SEM images of ensembles of MSs, while masses are measured in the trap for the MS used. Inferred unequal densities between MSs of different sizes is attributed to differences in the synthesis process.

| | Measurement configuration | | | | | |
|---|---|---|---|---|---|---|
| MS index | MS diameter [$\mu$m] | MS mass [pg] | MS dipole moment [e · $\mu$m] | MS position [$\mu$m] | MS dipole moment confinement plane | Integration time [$10^4$ s] |
| 1 | 7.56 | 312 | 182 | (9.6, 1.0, -1.3) | $yz$ | 4.5 |
| | | | | (9.6, 1.1, -2.9) | $yz$ | 5.4 |
| | | | | (10.3, 1.1, -3.6) | $yz$ | 4.5 |
| | | | | (10.2, 1.1, -3.1) | $yz^1$ | 4.95 |
| | | | | (10.3, 1.1, -3.0) | $xz^1$ | 4.95 |
| | | | | (9.2, 1.1, -3.0) | $xz^2$ | 4.95 |
| 2 | 9.98 | 543 | 9 | (14.7, 0.6, 0.2) | $yz$ | 10 |
| 3 | 9.98 | 1059 | 48 | (11.0, 1.1, -1.5) | $yz$ | 4.5 |
| | | | | (11.0, 1.1, -1.6) | $yz^3$ | 4.5 |
| | | | | (13.0, 1.1, -1.6) | $yz$ | 4.5 |

Table 1: Summary of measurement configurations used to produce the results in the main text. The MS position is given with the edge of the attractor defining $x = 0$, and the central gold finger on the attractor defining $y = 0$ and $z = 0$. In all cases, the measurement started with the MS spun up to an angular velocity $\sim 10^6$ rad/s. The ringdown time was separately verified to be long enough that the MS retained angular velocity $\gtrsim 5 \times 10^4$ rad/s at the end of the measurement. Each of the attractor and shield were independently electrically grounded for the duration of all measurements unless otherwise specified in the footnotes. The quoted values of the dipole moment are obtained for each MS in the trap.

---

[1] Attractor was DC-biased $-50$ mV relative to the electrically grounded shield.
[2] Attractor bias was sinusoidally modulated at 40.5 Hz, 150 mV$_\text{pp}$ relative to the electrically grounded shield.
[3] Attractor bias was sinusoidally modulated at 46.5 Hz, 200 mV$_\text{pp}$ relative to the electrically grounded shield.


\fi

\end{document}